# Highly anisotropic organometal halide perovskite nanowalls grown by Glancing Angle Deposition

*Javier Castillo-Seoane,[1,2] Lidia Contreras-Bernal,[1] Jose M. Obrero-Perez,[1] Xabier García-Casas,[1] Francisco Lorenzo-Lázaro,[1] Francisco J. Aparicio,[1,3] M. Carmen Lopez-Santos,[1,3] T. Cristina Rojas,[1] Juan A. Anta,[4] Ana Borrás,[1] Ángel Barranco,[1*] Juan R. Sánchez-Valencia[1,2 *]*

1    Institute of Materials Science of Seville (US-CSIC) Americo Vespucio 49, 41092 Seville (Spain); javier.castillo@icmse.csic.es; jmanuel.obrero@icmse.csic.es; lidia.contreras@icmse.csic.es; xabier.garcia@icmse.csic.es; franlorlaz@gmail.com; fjaparicio@icmse.csic.es; mclopez@icmse.csic.es; tcrojas@icmse.csic.es; anaisabel.borras@icmse.csic.es; angelbar@icmse.csic.es; jrsanchez@icmse.csic.es.

2    Atomic, Nuclear and Molecular Physics Department, Facultad de Física, University of Seville, Ad. Reina Mercedes s/n, 41012, Seville (Spain).

3    Department of Applied Physics I, University of Seville, Virgen de Africa 41011 Seville (Spain).

4.    Área de Química Física, Universidad Pablo de Olavide, Seville, 41013, (Spain) anta@upo.es

* jrsanchez@icmse.csic.es; angelbar@icmse.csic.es;

## Abstract

Polarizers are ubiquitous components in optoelectronic devices of daily use as displays, optical sensors or photographic cameras, among others. Yet the control over light polarization is an unresolved challenge as the main drawback of the current display technologies relays in significant optical losses. In such a context, organometal halide perovskites can play a decisive role given their flexible synthesis with under design optical properties such as a tunable bandgap and photoluminescence, and excellent light emission with a low nonradiative recombination rate. Therefore, along with their outstanding electrical properties have elevated hybrid perovskites as the material of choice in photovoltaics and optoelectronics. Among the different organometal halide perovskite nanostructures, nanowires and nanorods have lately arise as key players for the control of light polarization for lighting or detector applications. Herein, we will present the unprecedented fabrication of highly aligned and anisotropic methylammonium lead iodide (MAPI) perovskite nanowalls by Glancing Angle Deposition of $PbI_2$ under high vacuum followed by $CH_3NH_3I$ deposition at normal angle. Our approach offers a direct route for the fabrication of perovskite nanostructures virtually on any substrate, including on photovoltaic devices. The unparalleled alignment degree of the perovskite nanowalls provides the samples with strong anisotropic optical properties such as light absorption and photoluminescence, the latter with a maximum polarization ratio of $P$=0.43. Furthermore, the implementation of the MAPI nanowalls in photovoltaic devices provides them with a polarization-sensitive response, with a maximum photocurrent difference of 2.1 % when illuminating with the near-infrared range of the solar spectrum (>700 nm). Our facile vacuum-based approach embodies a milestone in the development of last generation polarization-sensitive perovskite-based optoelectronic devices such as lighting appliances or self-powered photodetectors.

## Introduction

In the last decade, organometal-halide perovskites (OMHP) have fascinated the research community due to their exceptional optoelectronic properties, finding applications as solar cells,[1,2] light emitting devices[3,4] and photodetectors.[5–7] Within these applications, the synthesis by vacuum deposition arises as an industrial scalable, low cost and environmentally friendly methodology to fabricate efficient, stable and durable optoelectronic devices.[8–11] On the other side, the anisotropic nanostructuration of OMHP such as nanorods, nanowires or nanoplatelets have been achieved by different routes,[6,12–14] which can be divided into template- and chemical-assisted growth:[15] the first make use of template structures such as electrospun fibers[16] or nanostructures such as pillars or grooves[17,18] to grow the OMHP in its interior, while the second, the most used, employ solution synthetic approaches to control the growth such as surfactants or anion-exchange reactions, among others.[12,19] One crucial characteristic of these semiconductor anisotropic nanostructures is their polarization-sensitive optoelectronic response.[15,20–22] Although many of our nowadays devices make use of polarizers to produce polarized light, there are several drawbacks such as the reduced intensity of the generated beam and/or their integration in micro and nanoscale devices, limiting the overall efficiency of the optoelectronic systems.[15,23]

The use of anisotropic OMHP nanostructures is proposed as one of the main candidates for the development of polarized light sources and absorbers due to their outstanding properties such as facile synthesis, tunable bandgap, narrow emission and high quantum photoluminescence yield.[12,24] Critically, this application requires the *ad hoc* alignment of the nanostructures, which is not straightforward in the standardized solution-based synthetic routes. Some alignment strategies can be found in literature such as evaporation-mediated, electric field, template and liquid crystal assisted, or oriented polymeric extruded or electrospun matrices.[25] These alternatives are strongly hampered by their difficult implementation and, especially, when dealing with large scales, limiting their applications in real optoelectronic devices.[15,23,25]

In this work, we propose for the first time, the application of an evolved version of the vacuum technique Glancing Angle Deposition (GLAD)[26–28] as an advanced alternative to the synthesis of anisotropic OMHP nanostructures. This approach provides the direct formation of the aligned nanostructures on most possible substrates, including photovoltaic devices.[29]

Our novel approach is a two-step fabrication procedure consisting of the room temperature deposition of PbI$_2$ at glancing angles (at 85º), followed by deposition of CH$_3$NH$_3$I at normal incidence (0º). Vacuum-sublimated methylammonium lead iodide (MAPI) perovskite layers are usually synthesized by co-evaporation of lead iodide and methylammonium iodide (CH$_3$NH$_3$I, MAI) precursors.[9,10] Although some examples can be found in the literature regarding the sequential vacuum evaporation at room temperature of both molecules,[30–32] the relatively compact initial inorganic film hinders the penetration of the methylammonium iodide along the entire bulk crystal to react and form the perovskite. With our methodology, highly porous PbI$_2$ layers are formed in the first instance, allowing the MAI to react with the entire structure. As a result, highly anisotropic MAPI nanostructures resembling "nanowalls" have been fabricated. Such an unparalleled alignment degree which is, to the best of our knowledge, the highest reported for GLAD, endows the samples with high anisotropic optical properties such as UV-Visible absorption and photoluminescence (PL). Moreover, their implementation in n-i-p OMHP solar cells can be used to develop self-powered polarization-sensitive photodetectors along the visible range. Therefore, the application of Glancing Angle Deposition provides highly anisotropic, microstructure and thickness controlled OMHP, with potential for large-area fabrication. This technique is fully compatible with microelectronic and optoelectronic processing methods including CMOS and roll-to-roll technologies, opening the path towards the development of tunable anisotropic optoelectronic devices based on OMHP.

**Results and Discussion**

PbI$_2$ was sublimated under high vacuum conditions on substrates positioned at normal incidence (0º) and glancing angles and then exposed to MAI precursor. Figure 1 shows the XRD peaks of PbI$_2$ normal (0º) deposited sample (red and green curves) and GLAD at 85º (blue and brown) before and after deposition of MAI. Whereas the compact $PbI_2$ sample deposited at 0º does not show any change after the deposition of the MAI in the XRD spectra (green curve), the Glancing Angle deposited one (at 85º) suffer an almost complete transformation to $MAPbI_3$ (brown curve), as can be observed by the loss of the XRD peak associated to $PbI_2$ (001) at 12.7º and appearance of the MAPI (110) one at 14.2º. The peaks associated with every crystalline (hkl) plane are highlighted with orange triangles ($PbI_2$) and brown squares (MAPI). It can be observed a high texturization in the diffractogram of MAPI, corresponding to a preferential orientation along the (110) crystalline plane, where no other peak different than (110), (220), (330) and (440) is visible. It is also shown a preferential texturization along the (001) direction in the GLAD $PbI_2$ before the exposition of MAI, although additional minor peaks associated to (100) and (110) at 22.5 and 39.5º, respectively, as well as oxidation products such as $PbO_2$ with a maximum at ca. 28.5º, are present. The Full Width at Half Maximum (FWHM) of the main XRD peak for GLAD samples, is much higher for $PbI_2$ (001) (0.45º) than for MAPI (110) (0.18º), indicating that the deposition of MAI (performed over $PbI_2$ GLAD samples at 50ºC), induces a preferential crystal transformation and growth as reported elsewhere.[33] In addition, the transformation to MAPI is appreciable by the naked eye, since it changes its colour from yellow to brown as it is shown in the insets of Figure 1.

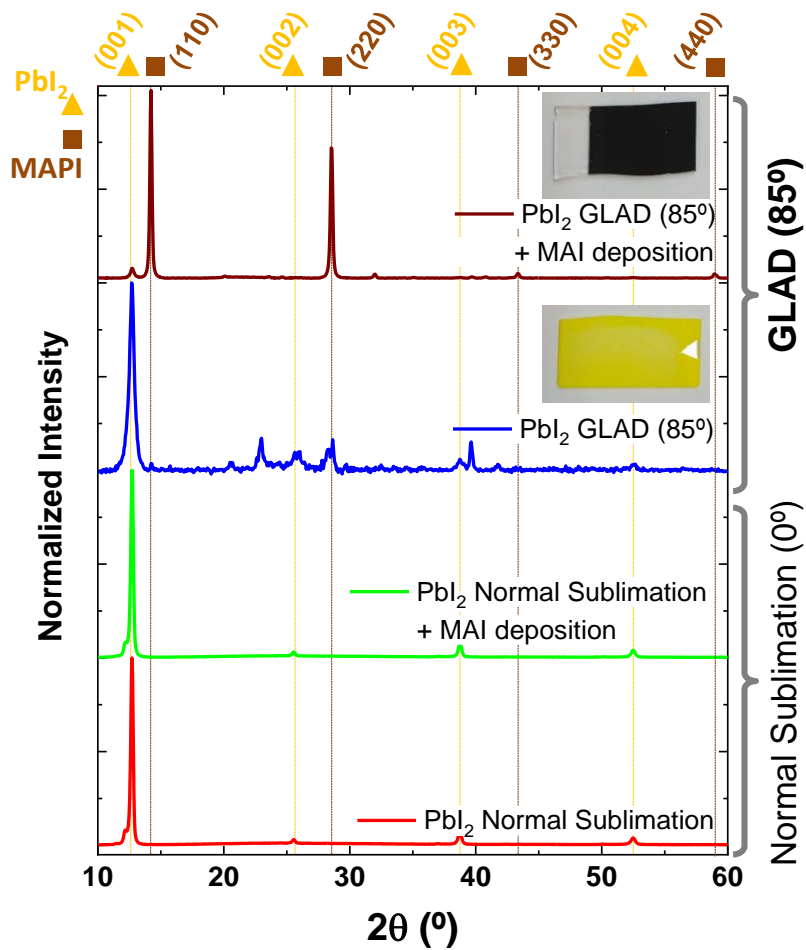

**Figure 1.** X-Ray Diffractograms of PbI$_2$ normal deposited sample (red and green curves) and GLAD at 85º (blue and brown) before (red and blue) and after (green and brown) deposition of MAI. The peaks associated with every crystalline (hkl) plane are indicated with orange triangles (PbI$_2$) and brown squares (MAPI). The insets show the corresponding image of PbI$_2$ GLAD before and after MAI deposition.

Figure 2 shows the top view SEM images at different magnifications of GLAD perovskite thin films deposited at 70º (a,b) and 85º (c,d) using the two-step sequential protocol detailed. GLAD samples present a highly porous structure that allows the MAI to diffuse along the PbI$_2$ layer and react to form the MAPI phase, in agreement with the transformation observed in the XRD pattern of Figure 1. It can be noted that the porosity is significantly enhanced for the highest deposition angle of 85º. The high magnification image of Figure 2 b,d) shows elongated nanostructures with widths (the size along y-direction) of around 120 and 80 nm (see also Figure S1 in Supplementary Information) and lengths (along x-direction) of ca. 1 and above 2 microns, for 70 and 85º of deposition angle, respectively. In both cases, highly oriented nanostructures that resemble nanowalls (NWs) can be observed along the x-axis, while leaving open void channels along the y-direction of ca. 500 nm for the 70º and of more than 1 micron for the 85º deposition angle. However, the films deposited at the highest glancing angle (85º) present the highest alignment along the x-axis and therefore the largest and more defined inter-wall spacing. Hence, we focused the study on this latter case.

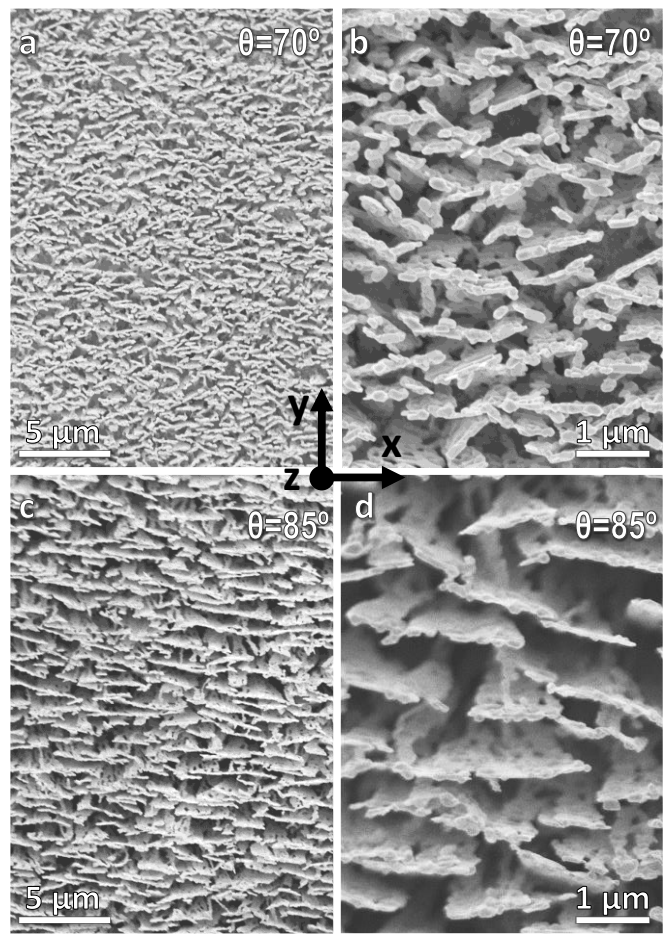

**Figure 2.** Top-view SEM images of GLAD perovskite thin films deposited at 70º (a,b) and 85º (c,d) at two different magnifications. Arrows are included to indicate the geometrical axes, where x corresponds to the aligment direction.

The characteristic cross-sectional SEM images of the 85º deposition angle sample are shown in Figure 3. As the elongated nanostructures are highly oriented along the x-axis, the cross-sectional views are different for samples cleaved along the y- or x-direction. Figure 3 a-b) and c-d) show these different cuts along the y-z and x-z planes, respectively. The microstructure observed in Figure 3 a) (y-z plane) consists of apparently one-dimensional nanowires slightly tilted toward the source of PbI$_2$ vapour during deposition. Contrarily the x-z plane views show a continuous layer with an apparent lower porosity.

The images confirm the growth of quasi vertical 2-Dimensional NWs with a strong orientation along the x-z plane. The coalescence of the nanostructures along the x-direction (perpendicular to the deposition direction) is commonly known as *bundling* and has been observed in many GLAD systems.[26,27,34,35] Unlike conventional *bundling* that is usually a nanocolumnar association, the nanostructures observed here present, to the best of our knowledge, the highest anisotropy observed for a GLAD nanostructure.[27]

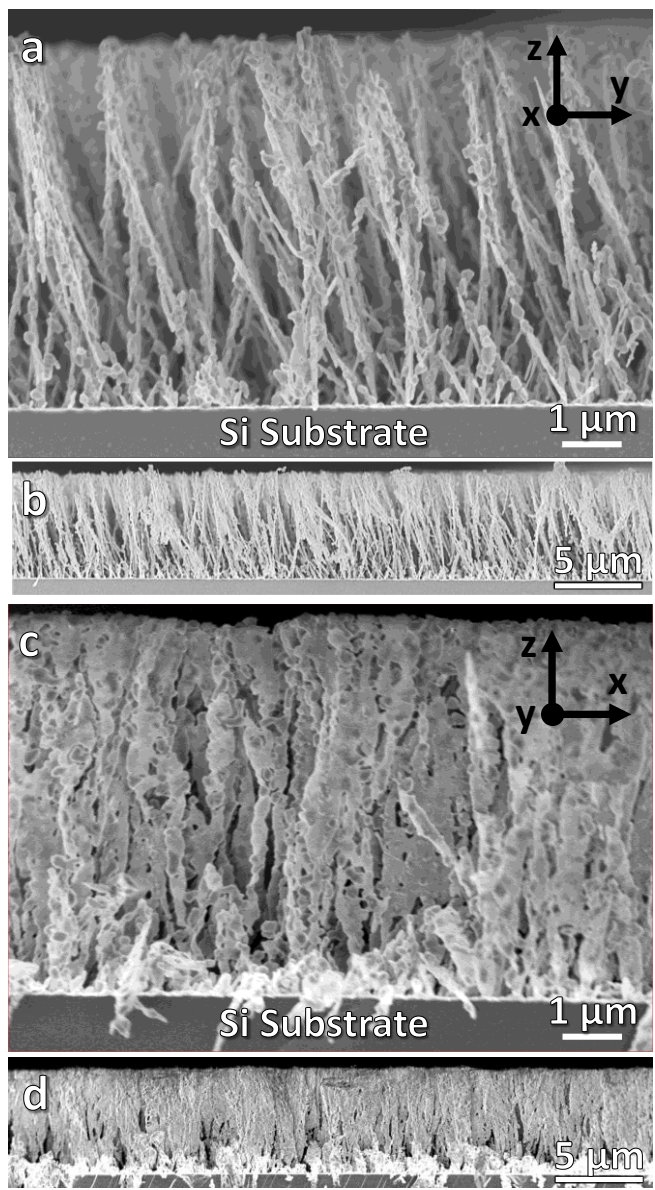

**Figure 3.** Cross-Sectional SEM images of GLAD perovskite thin films deposited at 85° cleaved along the y- (y-z plane, a-b) and x-direction (x-z plane, c-d) with different magnifications. Arrows are included to indicate the geometrical axes.

The TEM image of Figure 4e) shows an overview of transferred perovskite NWs (henceforward 85°), where many fragments (from the transfer procedure, see Experimental Section) with different sizes and shapes can be identified. Some of these fragments are stacked as it can be deduced from the darker appearance at some locations (highlighted with orange circles). A representative nanowall fragment is presented in Figure 4 a). One important observation is the very high sensitivity of the samples to the electron beam that degrades the fragments in a very short time. This effect is shown in Figure 4 a-b), where it can be noted how the original single perovskite nanowall (a) is degraded after few seconds under the 200 keV accelerated electron beam (b). The inset in Figure 4 b) is a magnified image of the area marked in red showing the damage induced by beam irradiation. The Selected Area Electron Diffraction (SAED) pattern of the Figure 4 b) (green circle) reveals a hexagonal arrangement that could be initially associated to the (111) surface of the tetragonal structure of MAPI. However, a close inspection to the SAED image shown in Figure 4 g) (obtained from the small fragment of Figure 4 f) reveals the transformation of the MAPI to $PbI_2$, as can be deduced from the distances extracted from the diffraction image. In particular, the pattern observed correspond to the [001] zone axis of the hexagonal crystal structure of $PbI_2$ (shown in Figure 4 d), where it can be identified the periodic distances of 2.2 and 3.9 Å, in good agreement with the (110) and (100) plane spacing.[36] Such damaging effect is typical in hybrid metal halide perovskites as has been previously reported elsewhere and has prompted the development of *ad hoc* tools to circumvent this issue such as controlling the electron dose over the specimen.[37,38] However, and despite this degradation to $PbI_2$, the results in Figure 4 allow us identifying a single crystalline growth all along the individual fragments observed. Thus, the SAED images of more than 10 ensembles show the same hexagonal arrangement (see Supplementary Information S2). These results are consistent with a highly texturized crystalline orientation of the MAPI nanowalls, supported by the strong (110) orientation observed in the XRD pattern of Figure 1. Although many articles have reported the increased or reduced texturization induced by Glancing Angle Deposition in systems such as metal, metal oxides or organic materials,[27,39–41] to the best of our knowledge this is the first time that such a high crystalline orientation is observed.

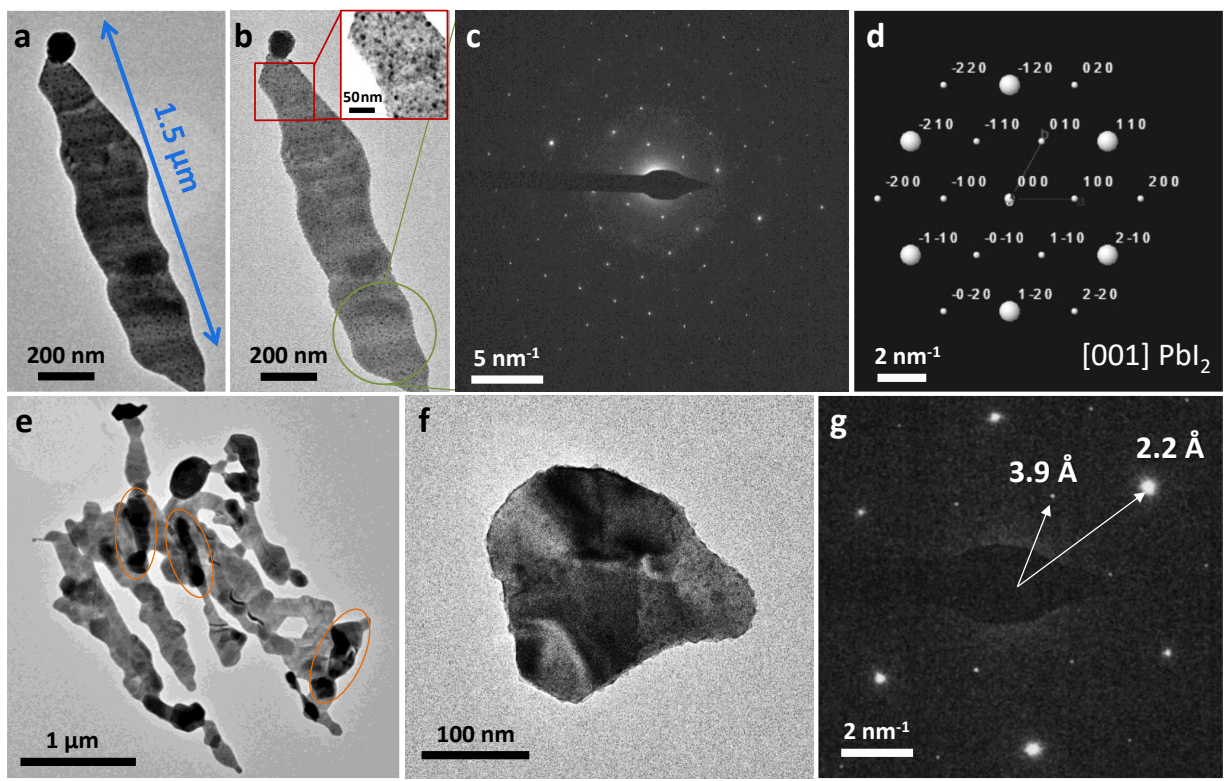

**Figure 4.** TEM images of individual fragments of MAPI NWs at short exposure to the e-beam (a) and after few seconds (b). The inset in b) shows a magnified image of the highlighted area. c) SAED of the green circle zone in b). e) Overview of the transferred samples. Stacked layers are highlighted with orange circles. f) MAPI fragment and g) corresponding SAED. The simulated diffraction pattern for the [001] axis zone of $PbI_2$ is shown in d).

As a direct consequence of the perovskite nanowalls alignment, the samples present strong anisotropic optical properties. Figure 5 a) shows the UV-Vis-NIR direct absorbance spectra using polarized light beams along the x- and y-directions (being the x-axis the alignment direction of the nanowalls as presented in Figure 2 c-d) of the sample deposited at 85°. It is worth stress herein that the proposed methodology allows a fine thickess control by simply increasing the deposition time without detriment in the composition homogeneity, being such one of the additional advantages over the chemical- and template-assisted approaches. Hence, since the 6 micrometer-thickness samples shown in Figures 2 and 3 present an almost complete absorption in the UV-Vis range for both polarizations, a sample with 2 µm thickness was used for the optical studies of Figure 5. To avoid degradation, the samples were measured in an inert atmosphere inside a homemade packing between two fused silica windows (see Experimental Section). Figure 5 b) presents the absorbance difference between x- and y-polarization and demonstrates the marked anisotropy shown in these samples, with a pronounced difference in the light absorbance between the two in-plane polarizations. It can be observed that the absorbance difference is almost zero for wavelengths below 450 nm. By contrast, the x- and y-polarized absorbance diverges above this wavelength (see Figure 5 a), depicting a maximum in the absorbance difference of ca. 1.5 at 800 nm.

Outstandingly, the photoluminescence (PL) of the MAPI aligned nanowalls (Figure 5 c) also reveals strong anisotropic properties, showing a much more intense PL emission polarized along the x-pol when excited with unpolarised light at 600 nm. Figures 5 d-f) show the polar diagram of the PL emission (illuminating with unpolarised excitation light) and excitation polarization (recording unpolarised emission light), using different excitation wavelengths, 400 (d), 500 (e) and 600 nm (f). First, it can be noted that both the anisotropy in the emission (black line) and excitation polarization (coloured line) increases with wavelength. The case of excitation at 600 nm (Figure 5 f) shows the typical lobe-shape pattern for both PL excitation and emission polarization. For shorter excitation wavelengths, the excitation polarization is nearly isotropic with a quasi-circular shape. By contrast, the emission polarization retains the lobe-shape pattern for all the wavelengths studied, with a continuous reduction in the anisotropy as the wavelength decreases.

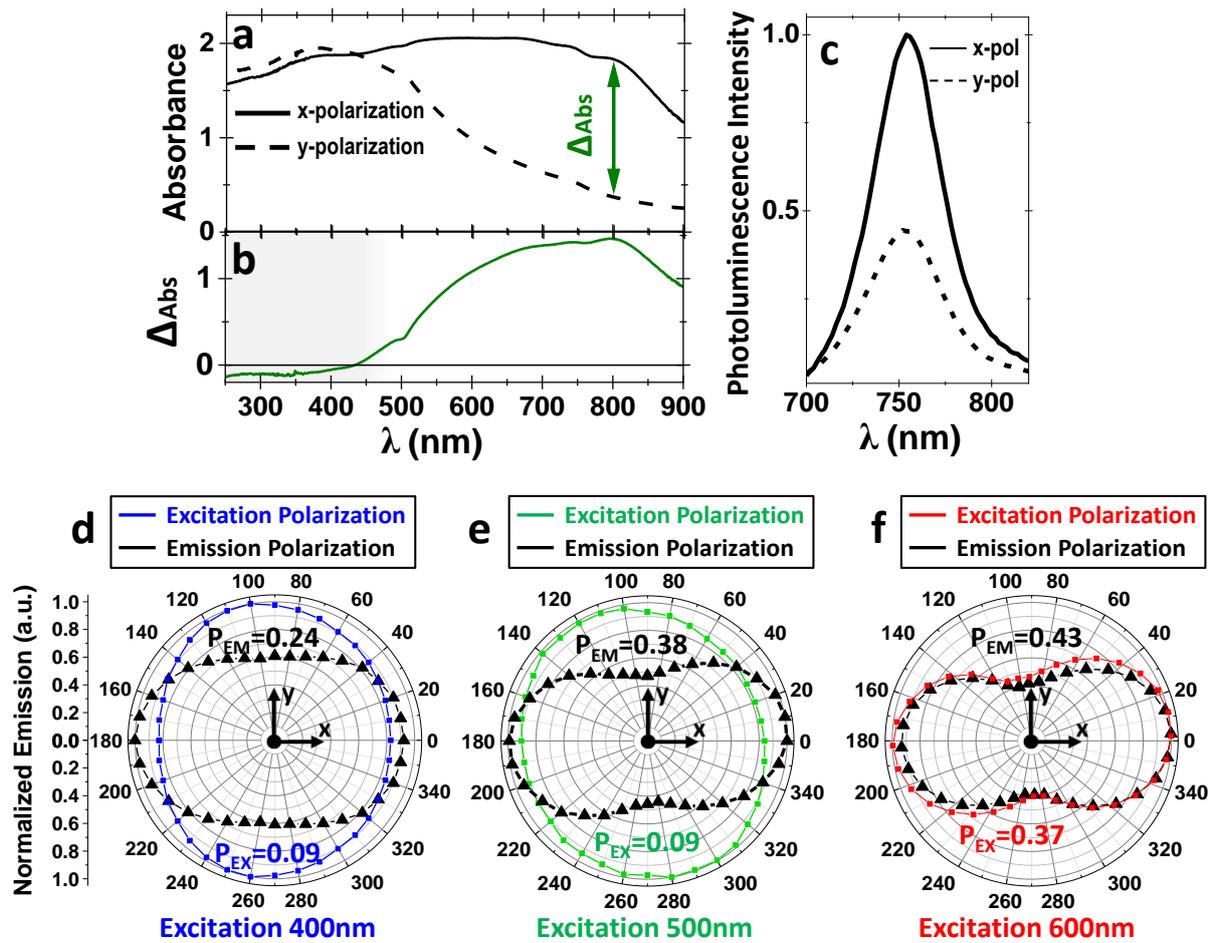

**Figure 5.** a) UV-Vis-NIR absorbance spectra of MAPI NWs using polarized light along x- (solid line) and y-polarization (dashed line) and b) Absorbance Difference ($\Delta_{Abs}$=Abs(x-pol) – Abs(y-pol)). c) Photoluminescence emission polarized spectra along the x- (solid line) and y-polarization (dashed line) exciting with unpolarised light at 600 nm. d-f) Azimuthal polar representation of the PL emission polarization (using unpolarised excitation light, black line) and excitation polarization (recording unpolarised emission light, coloured line), using 400 (d), 500 (e) and 600 nm (f) as excitation wavelengths.

From the PL polar representation obtained by exciting at 400 (d) and 500 nm (e), it can be noted that while the polarization ratio, *P*, (see S1.Electrostatic Approximation) for excitation polarization is very low ($P_{EX}$=0.09 for both illumination wavelengths), the emission polarization is significantly higher, $P_{EM}$=0.24 and 0.38 for 400 and 500 nm excitation wavelengths, respectively. This difference is consistent with the results reported by Täuber *et al.*, which characterize the PL excitation and emission polarization of individual MAPI nanowires, using a 488 nm laser as excitation source.[42] They observed a systematic difference between these excitation and emission polarization measurements for 28 different isolated nanocrystals at room temperature, with a much lower $P_{EX}$ value, in agreement with our measurements using a similar excitation wavelength (Figure 5e). Moreover, in the cited work[42] the maximum $P_{EM}$ for 28 individual nanocrystals is 0.37 while in our case is 0.38 for a macroscopic sample (probed over ca. 1 cm$^2$), i.e. the average all over our aligned anisotropic nanowalls. This value is highly remarkable, and one of the highest obtained for MAPI anisotropic nanostructures.[22,42] Furthermore, the $P_{EM}$ is even higher as the excitation wavelength increases, reaching 0.43 for excitation at 600 nm, closer to the bandgap of MAPI. These results could be initially attributed to the dielectric constant contrast between MAPI and the surrounding environment as detailed in S1, that should result in a polarization ratio of $P$=0.92. Although some authors have used this approximation for perovskite anisotropic nanostructures such as MAPbBr$_3$[22,43] or CsPbBr$_3$[44,45] with a certain agreement, the detailed study of individual MAPI nanowires[42] revealed that this approximation is not correct due to the strong differences observed in the excitation and emission polarization measurements, as observed here. This difference is attributed in literature to the orientation of the crystalline nanostructure: the room temperature tetragonal structure is non centrosymmetric and consequently the bond angles and lengths between Pb$^{2+}$ and I$^-$ differ for the two dissimilar planes (namely the equatorial and apical planes), producing modifications in the bandgap due to the hybridization of the 6s orbitals of lead and the 5p of iodide.[42] The smaller energy bandgap arises for the apical direction where the bond angles are smaller and possesses a transition dipole moment that results in a polarized photoluminescence. Consequently, it can be assumed that for polycrystalline ensembles, this effect is negligible. Surprisingly, these authors observed an isotropic behaviour in the excitation polarization by using excitation sources with energy well above the bandgap and foresaw a non-isotropic behaviour for lower energies, although it was not measured. This is the exact phenomenon accounted for our MAPI nanowalls and the first evidence that supports their hypothesis.[42] This result is also in agreement with the high texturization observed by XRD and TEM, indicating a preferential crystalline orientation all along the nanowalls.

As introduced above, the advantage of our approach is its granted full compatibility with the deposition over an ample variety of substrates. As a proof of concept, our MAPI NWs were implemented in a n-i-p solar cell (see Methods) with thicknesses of 350 and 500 nm to ensure a complete covering of the nanowalls by Spiro-OMeTAD films. Figure 6a presents a cross-sectional SEM micrograph of the device integrating MAPI nanowalls of 500 nm thickness, demonstrating their complete coverage. An additional thickness of 700 nm is also included, although it was verified that the tips of the nanowalls were not covered by Spiro-OMeTAD (See Figure S3).

As the MAPI NWs present strong anisotropic properties, the solar cell developed here can be used as a self-powered polarization photodetector. Figures 6 b-d) show the photocurrent increment of the different samples illuminated with x- and y-polarized light using the complete spectrum of the solar simulator (b) and blocking the illumination spectrum below 500 (c) and 700 nm (d) employing long-pass filters. To better observe the response, the illumination of the devices was periodically changed from x- to y-polarization every 10 seconds (as indicated in the graph with the grey background). In Figure 6 b) it can be noted that the device with 350 nm of MAPI (red line) does not show any significant dependence with the polarization. However, the MAPI nanowalls with 500 (blue) and 700 nm (green) thickness depict a certain photocurrent increase with polarized light, which is ca. 0.05 and 0.1 %. It is worth mentioning that this change is instantaneous for all the cases with a visible response. On the other hand, the illumination through the 500 nm long-pass filter (Figure 6c) produces a significantly higher polarization response, with a maximum photocurrent increase of ca. 0.9 % for the devices containing nanowalls of 350 and 500 thickness. Furthermore, the illumination with the long-pass filter at 700 nm (Figure 6 d), produces similar responses but significantly enhanced, reaching a maximum of 2.1 % of photocurrent increase for the device with 500 nm thickness. It can be noted that although the maximum photocurrent anisotropy is expected for the 700 nm sample, the NWs were not totally covered by the Spiro-OMeTAD layer and we observed a much faster degradation, which can also contribute to the reduction in the anisotropy observed.

As a first approximation, the anisotropic response could be related to the absorbance difference between x- and y-polarization, but in the case of 500 nm MAPI NWs this difference is very small, as can be seen in Figure S4. Additionally, by means of the 500 nm long-pass filter and even more with the 700 nm one, the polarization response is maximized. However, the absorbance difference is negligible above 700 nm (Figure S4), indicating a different mechanism for the anisotropic photocurrent response. As it happened with the photoluminescence excitation polarization of Figure 5d-f, the anisotropic response is lost for wavelengths below 600 nm and consequently the visible range where the excitation with polarized light contributes to the anisotropic photocurrents is expected to be redshifted. Thus, our results support even further the previous hypothesis about the crystalline-related anisotropy and represent for the first time the highly texturized growth of aligned MAPI nanostructures with strong anisotropic properties and their implementation in real devices such as the self-powered polarization photodetector presented.

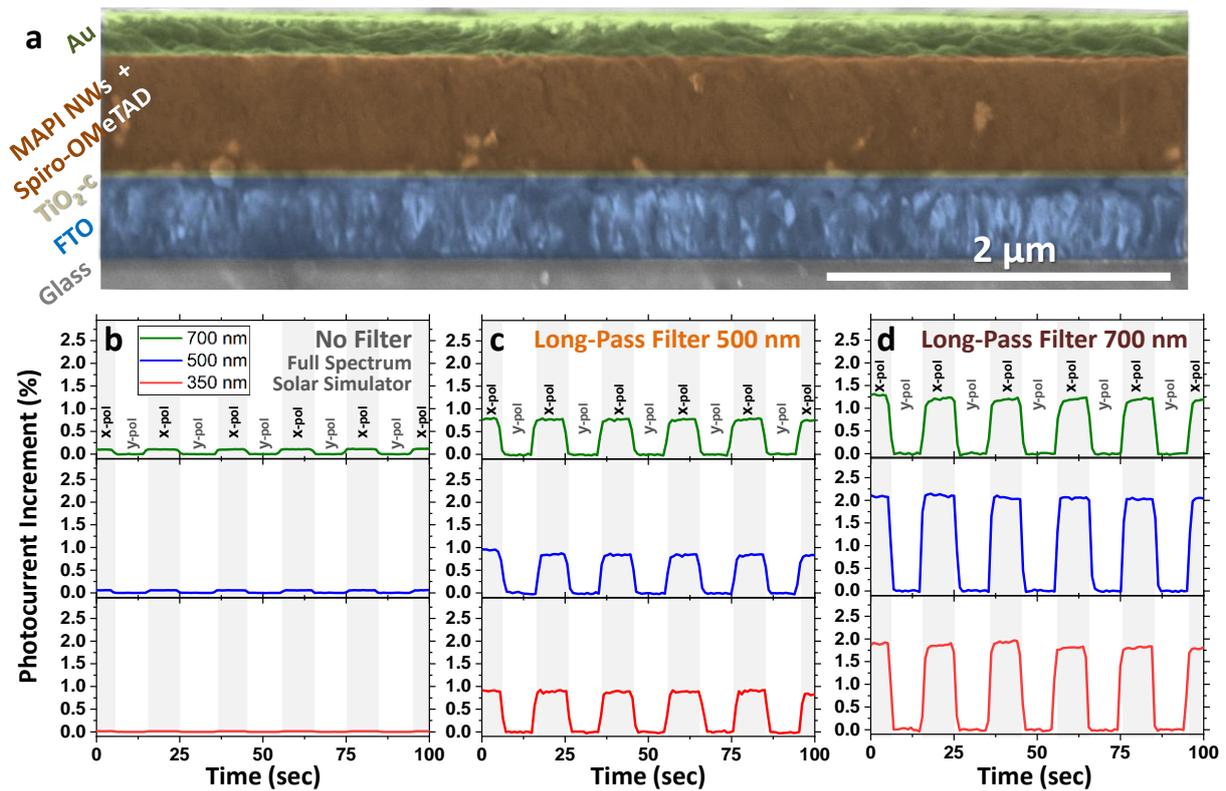

**Figure 6.** a) Cross-Sectional SEM image of the complete solar cell, implementing the MAPI NWs with 500 nm thickness. b-d) Photocurrent increment response recorded by changing between x- (higher photocurrents) and y-polarization (lower photocurrent) every 10 seconds for devices implementing MAPI NWs of 350 (red), 500 (blue) and 700 nm (green) thickness. The grey background indicates illumination periods with x-polarized light as indicated. The device was tested under illumination with the full solar spectrum (b) and through long pass filters at 500 (c) and 700 nm (d).

## Conclusions

Our work details the unprecedented synthesis of methylammonium lead iodide perovskite nanostructures by Glancing Angle Deposition, producing highly anisotropic nanowalls. At the highest angle tested, 85º, highly aligned nanowalls along one single axis (x-direction) of around 2-4 µm length and widths below 100 nm were synthesized. This anisotropic growth is related to the *bundling* effect, which is inherent to the GLAD technique. Unlike conventional *bundling*, the nanostructures observed present, to the best of our knowledge, the highest anisotropy observed for a GLAD nanostructure.[27] In addition, the XRD analysis reveals a high crystalline texturization along (110) plane for MAPI NWs, in agreement with the hexagonal arrangement observed in the SAED pattern and indicating that the approach can be used to direct the crystalline growth of MAPI.

The highly aligned nanostructures present a strong in-plane optical anisotropy, with a much higher absorbance for polarized light along the NW axis. The absorbance difference is negligible in the range of the spectrum below 450 nm and increases above this wavelength, reaching a value as high as $\Delta_{Abs}$=1.45 at 800 nm. Moreover, the photoluminescence also shows a wavelength-dependent anisotropic response with strong differences in the emission (using unpolarized excitation) and excitation polarization (using unpolarized emission). While the excitation at shorter wavelengths produces a nearly isotropic excitation polarization, the excitation at 600 nm yields a typical anisotropic lobe-shape pattern. At the same time, the emission polarization is anisotropic for all the wavelengths studied and has been quantified with the polarization ratio which magnitude increases with wavelength and reaches values as high as 0.43. This result markes a milestone in the state of the art, especially taking into account that only isolated nanorods have so far provided values close to this, and here we have a macroscopic sample composed of aligned individual anisotropic nanowalls.

Our approach produces nanostructures directly supported on almost any substrate since it takes place at low temperatures and the only requirement is their flatness. As proof of this, the MAPI nanowalls were implemented in a solar cell used as a self-powered polarization photodetector. The excitation with polarized light produces different photocurrents, being higher for the direction of the nanowalls. The response was monitored by illuminating with a solar simulator using the full spectrum, and blocking a part of it by long-pass filters. In this latter way, the photocurrent response was maximized, especially at longer wavelengths, with a 2.1 % photocurrent increase between x- and y-polarization.

The responses of the NWs against the excitation with polarized light, i.e. the PL excitation and photocurrent polarization, have been discussed in the context of the recent literature regarding photoluminescence of individual MAPI nanowires.[42] The authors explained the differences between the photoluminescence excitation and emission polarization due to the crystalline orientation and foresaw that they should disappear by illuminating with longer wavelengths. This prediction has been verified in our work, obtaining a high PL anisotropy by using 600 nm excitation wavelength. A similar wavelength dependence has been observed in the NWs implemented in a solar cell, since the illumination with polarized light at short and long wavelengths produces an isotropic and anisotropic photocurrent response, respectively. Even though its experimental simplicity, the application of GLAD has already revolutionized the fabrication of tailored porous and sculptured films in topics ranging from energy to biomaterials, including optics and optoelectronics. We have demonstrated here the potential application of GLAD for the synthesis of highly anisotropic OMHP nanostructures with many operational advantages: i) direct integration and alignment degree of the NWs in virtually any substrate; ii) environmental friendly methodology since it is a solventless process at low temperature and iii) compatibility with industrially scalable processes. It is worth mentioning that the generality of our methodology will allow the synthesis of other OMHP, for instance, with photoluminescence and photocurrent responses shifted to the visible. The only additional requirement for the growth of these NWs with respect to conventional vacuum sublimation is the tilting of the substrates. For these reasons, we believe that our results will increase the potential of the development of full vacuum-processed polarization-sensitive optoelectronic devices.

## Experimental Section

*Sequential Glancing Angle Deposition of MAPI*: Lead Iodide ($PbI_2$) and methylammonium iodide (MAI) were acquired from TCI and used as received. Both precursors were sublimated by means of two separated Knudsen cells and heated up to temperatures around 330 and 160 ºC for $PbI_2$ and MAI, respectively. The initial $PbI_2$ deposition was carried out at pressures below $5\times10^{-6}$ mbar on fused silica and Si(100) wafer pieces. The deposition was also performed over FTO comercial pieces as detailed in the next subsection. The substrates were positioned at oblique and glancing angles of 70 and 85º with respect to the evaporation source by means of a transfer bar that allows both the angle control by the use of a goniometer and to transfer them into a glovebox. The deposition rate was controlled by means of a quartz crystal monitor (QCM) with a constant growth rate of 1.5 nm/s. After this step, the MAI was sublimated onto the $PbI_2$ deposited samples placed at 0º with respect to the evaporation source. This was performed with the samples at a constant temperature of 50 ºC. We observed that the deposition of MAI could not be properly controlled by the QCM (the initial growth rate achieved during the first minutes could not be maintained for the whole deposition). We also noted that differently from the $PbI_2$ deposition, the MAI sublimation produces and increase in the pressure of the system. Based on these observations, the final protocol used for MAI deposition was to heat the MAI to a temperature of 160 ºC, that produces an increase in the pressure to around $5\times10^{-4}$ mbar. Then, we reduced the pumping flux of the turbomolecular pump by means of a butterfly valve and fixing the pressure to $10^{-3}$ mbar. With this protocol we obtained reproducible MAPI NWs samples.

*Photodetector multilayer*: MAPI NWs-containing devices were fabricated on fluorine-doped tin oxide (FTO) glasses (TEC 15 Pilkington, resistance 15Ω/square, 82-84.5% visible transmittance) patterned by laser etching supplied by XOP Glass. FTOs were cleaned by ultrasonic bath following the solvent sequence: 1. Hellmanex water solution (2:98 vol/vol ratio); 2. Deionized water; 3. Isopropanol; 4. Acetone. A UV ozone treatment for 15 minutes was done as a final cleaning step. After that, 50 nm of compact $TiO_2$ was deposited on top of the FTO substrates by spray pyrolysis. The sprayed solution comprised 1 ml of titanium diisopropoxide bis(acetylacetonate) solution (75% in 2-propanol, Sigma-Aldrich) and 14 ml of absolute ethanol. The solution was sprayed on heated substrates (450 ºC) using pure oxygen as a carrying gas. The samples were kept at 450 °C for 30 min for the formation of the TiO$_2$ anatase phase. The compact TiO$_2$ layer was doped with lithium. For that, a lithium solution precursor was prepared with 10 mg of bis(trifluoromethane)sulfonimide lithium salt (Sigma-Aldrich) in 1 ml of acetonitrile (Across Organics). Then, this solution was spin coated on compact layer at 3000 rpm and 2000 rpm/s for 10 seconds. The samples were quickly again annealed at 450 ºC for 30 min (using reported temperature ramp).[46] Once the electrodes were cooled down, these were moved into a glove box to deposit a thin (30 nm) methylammonium lead iodide (MAPI) perovskite layer. First, a MAPI precursor solution was obtained using dimethylformamide (DMF, Acros) as solvent in which, first, dimethyl sulfoxide (DMSO, Acros), then PbI$_2$ (heated up to 70ºC without stirring) and finally, MAI (at room temperature) were dissolved. The concentration of the three compounds, DMSO, PbI$_2$ and MAI, was 1.3 M. Finally, this precursor solution is diluted (1:4 vol%) using a DMF:DMSO solvent-mix (9:1 vol%). Then, MAPI film was deposited on lithium-doped TiO$_2$ substrate by spin-coating following a reported one-step protocol: 5000 rpm for 50 seconds. Six seconds after the beginning of the spinning, the DMF was selectively washed with nonpolar diethyl ether (Sigma-Aldrich). Afterward, the samples were annealed at 65 ºC for 1 min and then for 2 min at 100 °C. On top of these samples, sequential Glancing Angle deposited MAPI NWs were synthesized, except for the reference photodetector samples. The samples were transferred to the vacuum system with the aid of a transfer bar and a gate valve coupled to the glovebox. After the deposition of the MAPI NWs, the samples were transferred again to the glovebox (without exposing them to the air), and Spiro-OMeTAD was spin coated on top of them in a two-step protocol: 1) 500 rpm for 20 seconds; 2) 1000 rpm for 10 seconds. The hole transport material solution was obtained from 72.3 mg of Spiro-OMeTAD (Merck) dissolved in 1 ml of chlorobenzene (Acros). The Spiro-OMeTAD solution was doped with 6.6µl of a tris(2-(1H-pyrazol-1-yl)-4-tert-butylpyridine)cobalt(III) (FK209, Sigma-Aldrich) stock solution (400 mg in 1 ml of acetonitrile), 14.5 µl of

tri[bis(trifluoromethane)sulfonimide]lithium salt (LiTFSI) stock solution (520 mg of LiTFSI in 1 ml of acetonitrile) and 26µl of 4-tert-butylpyridine (tBP, Sigma-Aldrich).

Finally, 60 nm of gold was deposited by thermal evaporation under vacuum at pressures between $1\cdot 10^{-6}$ and $1\cdot 10^{-5}$ mbar on a water-cooled sample holder at a growth rate of 0.1 Å/s the first 20 nm and then 1 Å/s.

*Characterization of films and photodetector:*

Scanning Electron Microscopy (SEM) images were acquired in a Hitachi S4800, provided with a field emission gun and using electrons accelerated to 2 kV. The samples were deposited on Si(100) wafers, that allows a correct cleavage to observe the cross-section images. The samples were mounted on the SEM sample-holder inside the glovebox and packed in inert atmosphere for their transportation to the microscope. After that, the samples were exposed to the atmosphere a few seconds for their transfer to the SEM.

Transmission Electron Microscopy (TEM) images and Selected Area Electron Diffraction pattern were obtained using a Scanning - TEM microscope, TALOS F200S from FEI company, working at 200 kV with 0.25 nm resolution. To perform these measurements, the samples were scratched on top of a holey Carbon TEM copper grid. This procedure was performed inside a glovebox and the grids were packed under inert atmosphere for their transportation. The grids were exposed to air for a very short time during their positioning on the TEM sample holder. The program EjeZ from the University of Cadiz[47] were used to simulate the digital diffraction patterns.

Absorbance spectra were acquired in a Varian Cary-100 Conc UV-Vis spectrometer in the range 200-1000 nm by placing a Glan-Taylor polarizer in the optical path of the light beam. Fluorescence spectra were recorded in a Jobin Yvon Fluorolog-3 spectrofluorometer using the front face configuration. The equipment is provided with two automated polarizers after the excitation and emission monochromators, respectively. To correct the polarization dependence of the detector with the polarization of the light, a MAPI layer prepared by solution was used as isotropic emitter. In order to avoid degradation, samples were sealed in inert atmosphere inside a homemade packing between two fused silica windows, for both measurements, absorbance and photoluminescence.

For structural characterization, X-ray diffractograms were acquired in a Panalytical X'PERT PRO instrument in the Bragg-Brentano configuration using the Cu $K_\alpha$ (1.5418 Å) radiation as the excitation source.

Short circuit Current curves versus time were measured under a solar simulator (ABET-Sun2000) with an AM 1.5G filter. The samples were measured in a closed holder under $N_2$ atmosphere. A depolarizer provided by Varian was used to reduce the partial polarization of the solar simulator. After this, we used a Glan-Taylor polarizer (Varian) that was periodically rotated from 0 to 90º every 10 seconds and below, a long pass filter of 500 and 700 nm provided by Edmund-Optics. The reference sample, which consists of a thin layer of MAPI synthesized by spin coating, also shows a certain dependence with the illumination polarization. This effect, attributed to the partial polarization of the light source (the solar simulator), could not be totally suppressed employing a depolarizer. In order to correct the photocurrent increment of the samples with respect to the reference, the curves were initially normalized (from 0 to 1) and subtracted a baseline. After that, the contribution to the photocurrent increment of the reference samples were subtracted to the MAPI NWs containing photodetectors.

## Supplementary Information

Supplementary Information is available.

**Acknowledgements**

We thank the AEI-MICINN (PID2019-110430GB-C21, PID2019-110430GB-C22 and PID2019-109603RA-I0), the Consejería de Economía, Conocimiento, Empresas y Universidad de la Junta de Andalucía (PAIDI-2020 through projects US-1263142, ref. AT17-6079, P18-RT-3480), and the EU through cohesion fund and FEDER 2014–2020 programs for financial support. CLS and JS-V thank the University of Seville through the VI PPIT-US. JS-V and Lidia Contreras acknowledge the Ramon y Cajal and Juan de la Cierva Spanish National programs, respectively. FJA also thanks to the EMERGIA Junta de Andalucía program. The projects leading to this article have received funding from the EU H2020 program under the grant agreements 851929 (ERC Starting Grant 3DScavengers). JAA thanks AEI-MICINN for SCALEUP SOLAR-ERA.net project PCI2019-111839-2

Supplementary Information

# Highly anisotropic organometal halide perovskite nanowalls grown by Glancing Angle Deposition

*Javier Castillo-Seoane,[1,2] Lidia Contreras-Bernal,[1] Jose M. Obrero-Perez,[1] Xabier García-Casas,[1] Francisco Lorenzo-Lázaro,[1] Francisco J. Aparicio,[1,3] M. Carmen Lopez-Santos,[1,3] T. Cristina Rojas,[1] Juan A. Anta,[4] Ana Borrás,[1] Ángel Barranco,[1*] Juan R. Sánchez-Valencia[1,2] ]*

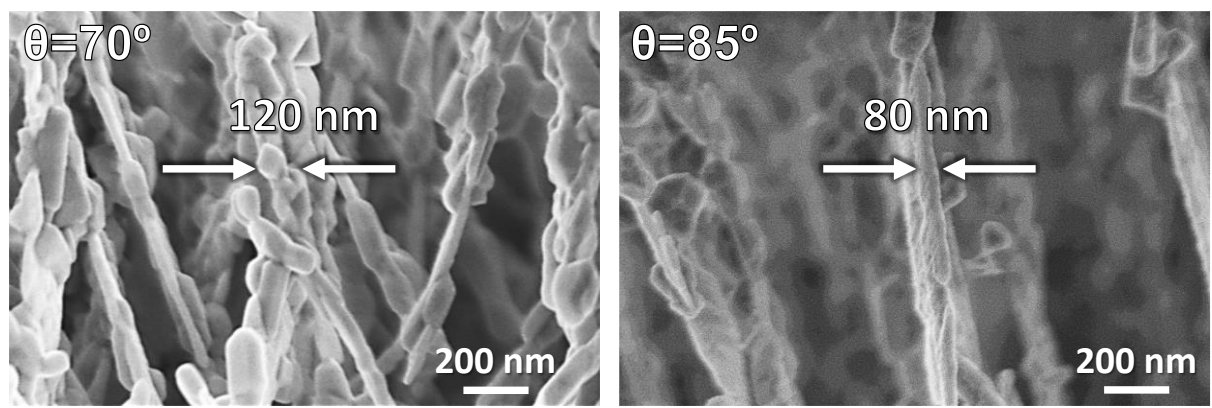

**Figure S1.** High magnification cross section SEM image of GLAD perovskite thin films deposited at 70º (left) and 85º (right) cleaved along the y-z plane

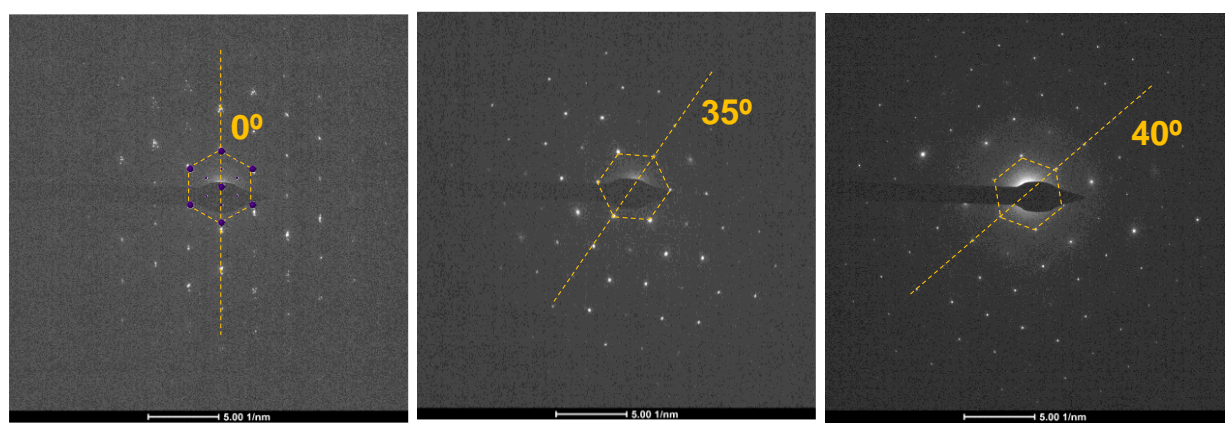

**Figure S2.** Selected Area Electron Difraction (SAED) images of representative MAPI NWs ensembles. The orange dashed line indicate the rotation angle with respect to the vertical, allowing the identification of the same hexagonal arrangement.

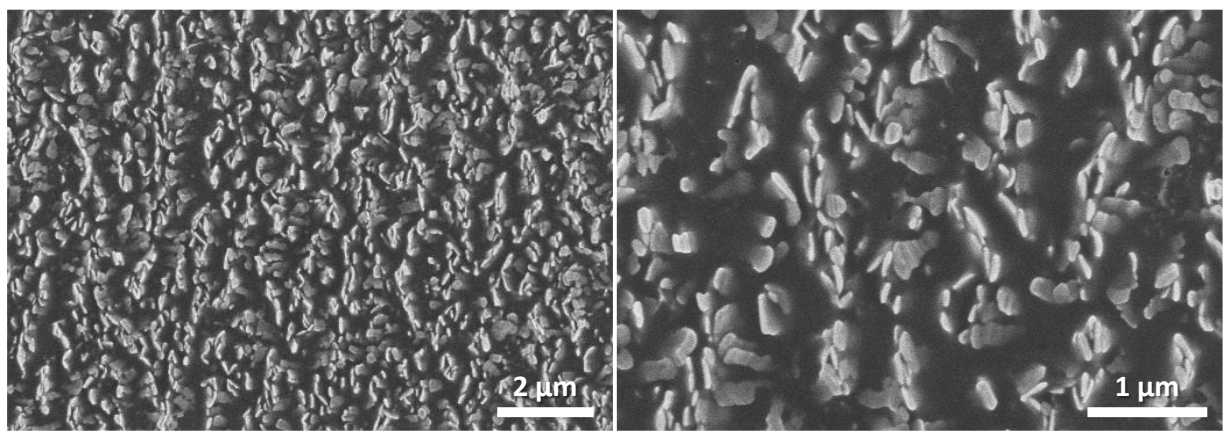

**Figure S3.** Top-view SEM image of the 700 nm thickness MAPI NWs sample covered by the Spiro-OMeTAD layer at different magnifications. It can be noted that the tip of the NWs protrude from the surface of the Spiro-OMeTAD layer

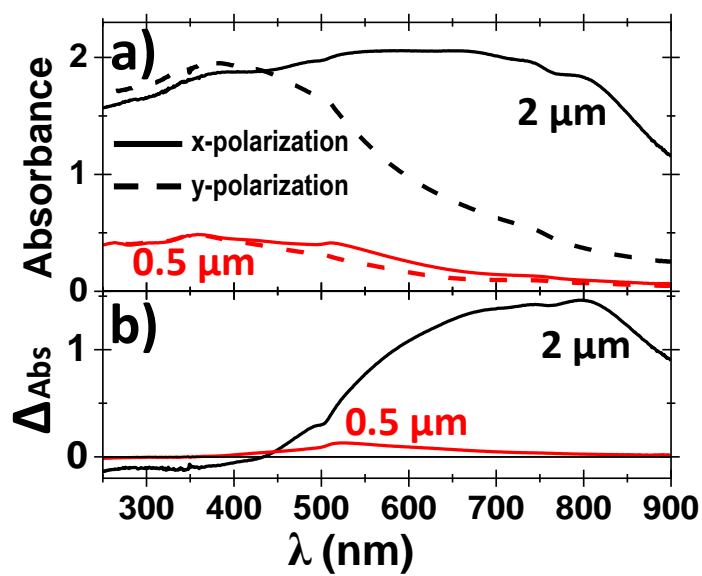

**Figure S4.** a) UV-Vis-NIR absorbance spectra of MAPI NWs of 2 µm (black) and 500 nm (red line) using polarized light along x- (solid line) and y-polarization (dashed line). b) Absorbance Difference ($\Delta_{Abs}$=Abs(x-pol) – Abs(y-pol)) for these two samples.

**S1. Electrostatic approximation.**

A useful magnitude to quantify the photoluminescence anisotropy is the polarization ratio ($P$), defined as: $P = \frac{I_{Max} - I_{Min}}{I_{Max} + I_{Min}}$, where, $I_{Max}$ and $I_{Min}$ are the maximum and minimum PL intensity and correspond to the parallel $I_{\parallel}$ (x-direction) and perpendicular axis $I_{\perp}$ (y-direction) to the nanowalls.

The photoluminescence anisotropy reported for the MAPI nanowalls could be initially attributed to the dielectric constant contrast between MAPI ($\varepsilon$) and the surrounding environment ($\varepsilon_0$). While the electric field along the x-axis (parallel to the nanowall) is not strongly affected by the nanostructuration, the field along the y-axis is attenuated due to the contrast of the dielectric constants between the perovskite and air. According to this electrostatic approximation, the excitation electric field coupled into the sample along y-axis can be expressed as:

$$E_i(y-axis) = \frac{2\varepsilon_0}{(\varepsilon+\varepsilon_0)} E_0,$$

where $\varepsilon$ and $\varepsilon_0$ are the dielectric constants of MAPI (ca. 9 at 600 nm)[48] and vacuum, respectively. Meanwhile, along the x-axis, $E_i$ is nearly equal to the incoming electric field:[44]

$$E_i(x-axis) = E_0$$

The photoluminescence intensity relates to the electric field by:

$$I_{Max} = I_{\parallel} = E_0^2 \quad \text{and} \quad I_{Min} = I_{\perp} = \left(\frac{2\varepsilon_0}{(\varepsilon+\varepsilon_0)}\right)^2 E_0^2$$

Therefore, the polarization ratio can be calculated as:

$$P = \frac{1 - \left(\frac{2\varepsilon_0}{(\varepsilon+\varepsilon_0)}\right)^2}{1 + \left(\frac{2\varepsilon_0}{(\varepsilon+\varepsilon_0)}\right)^2} = \mathbf{0.92}$$